\documentclass[showpacs,preprintnumbers,amsmath,amssymb]{revtex4}

\usepackage[usenames,dvipsnames]{color}
\usepackage{graphicx}
\usepackage{dcolumn}
\usepackage{bm}
\newcommand{\me}{\mathrm{e}}%
\newcommand{\mi}{\mathrm{i}}%
\newcommand{\dif}{\mathrm{d}}%
\usepackage{ulem}

\begin{document}

\title{Revisiting $1^{-+}$ and $0^{++}$ light hybrids from Monte-Carlo based QCD sum rules}

\author{Zhu-feng Zhang$^1$, Hong-ying Jin$^2$ and T. G.  Steele$^3$\\
$^1$Physics Department, Ningbo University, Zhejiang Province, P. R. China\\
$^2$Zhejiang Institute of Modern Physics, Zhejiang University, Zhejiang Province, P. R. China\\
$^3$ Department of Physics and Engineering Physics,
University of Saskatchewan, Saskatoon, Saskatchewan, Canada S7N 5E2}

\date{\today}

\begin{abstract}
In this paper, we re-analyze the $1^{-+}$ and $0^{++}$ light hybrids from QCD sum rules with a Monte-Carlo
based uncertainty analysis. With 30\% uncertainties in the accepted central values for QCD condensates and other input parameters,
we obtain a prediction on $1^{-+}$ hybrid mass of $1.71 \pm 0.22$\,GeV, which covers the mass of $\pi_1(1600)$.
However, the  $0^{++}$ hybrid mass prediction is more than 4\,GeV, which is far away from
any known $a_0$ meson. We also study the correlations between the input and output parameters of QCD sum
rules.
\end{abstract}

\pacs{12.38.Lg, 12.39.Mk, 14.40.Rt}
\maketitle

\section{Introduction}

There are several mesons with exotic quantum numbers (e.g., $1^{-+}$) listed in the Review of Particle
Physics \cite{Beringer:1900zz}. It is  appropriate to pay special attention on these mesons because they
cannot be accommodated in the conventional quark model.
In recent years, studies of their possible structures have become an important focus in hadronic physics.
For $1^{-+}$ mesons,  two possible structures have been studied in the literature:
 four-quark states \cite{Zhang:2001sb,Zhang:2004nb,General:2007bk,Chen:2008qw}
and hybrid states \cite{Balitsky:1982ps,Govaerts:1983ka,Latorre:1984kc,Barnes:1995hc,Bernard:1997ib}
(See Ref.~\cite{Narison:2009vj} for a recent review on studies of the $1^{-+}$ state).

The possible hybrid structures of $1^{-+}$ mesons have been studied by using many different methods,
including flux-tube model, lattice QCD and QCD sum rules.
The QCD sum rules method (QCDSR) is a very important nonperturbative method in hadronic physics. Since it was introduced
by Shifman et al. \cite{Shifman:1978bx,Shifman:1978by}, QCD sum rules have given numerous predictions on
hadron properties \cite{book2}. However, it has also been argued that there are
some shortcomings in traditional QCD sum rule analysis methodologies (see \cite{Leinweber:1995fn} for a detailed discussion). Mainly,
the continuum threshold $s_0$ could not be completely constrained  in the framework of
traditional QCD sum rules, which causes a significant uncertainty.
This shortcoming is reflected in many works, especially  in the case of $\Theta^+(1540)$, where many QCDSR works
gave the ``right" predictions \cite{Zhu:2003ba,Sugiyama:2003zk},
however, the signals of $\Theta^+(1540)$ are in fact statistical fluctuation effects rather
than true resonances \cite{Beringer:1900zz}.
To overcoming these shortcomings, Leinweber introduced a new Monte-Carlo based uncertainty analysis
into QCD sum rules and re-studied the $\rho$ meson and nucleon
 \cite{Leinweber:1995fn}.
Later, this procedure was used for predicting the decuplet baryon spectrum \cite{Lee:1997ix}, nucleon axial vector
coupling constants \cite{Lee:1996dc} and hadron magnetic moments \cite{Lee:1997ne,Wang:2008vg}. All these
studies are fruitful. Recently, a mathematica package MathQCDSR was provided \cite{Wang:2011zzx} to facilitate the uncertainty analysis.

In this paper,  we restudy the masses of $1^{-+}$ and $0^{++}$ light hybrids
by using QCD sum rules with Leinweber's Monte-Carlo based uncertainty analysis.
We first provide the operator product expansion results for correlators of the relevant
light hybrid currents followed by a brief introduction to the Monte-Carlo based QCD sum rules analysis. After that,
the masses for $1^{-+}$ and $0^{++}$ light hybrids are discussed respectively. Finally, we give
a summary and some further discussion.

\section{The operator product expansion of the correlator}

The current $j_\mu=\bar q\gamma_\nu \mi g G^a_{\mu\nu}T^aq$ can annihilate both $1^{-+}$ and $0^{++}$
states. To predict the properties of these associated states in QCD sum rules, the  relevant two-point correlator is necessary. In the present
case, it can be written as
\begin{equation}
\Pi_{\mu\nu}(q^2)=\mi\int \dif^4 x \me^{\mi q x}\langle 0|T j_\mu^{\textrm{ren}}(x) j_\nu^{\textrm{ren}\dagger}(0)|0\rangle
=(q_\mu q_\nu -g_{\mu\nu} q^2) \Pi_v(q^2)+q_\mu q_\nu\Pi_s(q^2),
\end{equation}
where the current $j_\mu^{\textrm{ren}}=(1+\frac{2}{9}\frac{g^2}{\pi^2}\frac{1}{\varepsilon}) j_\mu$ is the renormalized
current. In the QCD sum rules formulation, we focus on the operator product expand (OPE) of the
invariant correlators  $\Pi_v$ and $\Pi_s$, which correspond to the
contributions from $1^{-+}$ and $0^{++}$ states respectively. After
Borel transformation, they can be written as \cite{Balitsky:1982ps,Govaerts:1983ka,Govaerts:1984bk,
Latorre:1984kc,Latorre:1985tg,Balitsky:1986hf,Chetyrkin:2000tj,Jin:2000ek,Jin:2002rw}
\begin{equation}
\Pi^{\textrm{OPE}}_{v/s}(\tau)= a_{11}\frac{-2}{\tau^3} +a_{12}\frac{2}{\tau^3} (2 \gamma_E-3+2\ln(\tau \mu^2))
+b_{11}\frac{-1}{\tau}+b_{12}\frac{2}{\tau}(\gamma_E+\ln(\tau\mu^2))+c_{11}+c_{12}(-\gamma_E-\ln(\tau\mu^2))
+d_{11}\tau,
\label{sum_rule}
\end{equation}
where $\gamma_E$ is the Euler's constant,  $\tau$ is the Borel-transform parameter and we use $\Pi^{\textrm{OPE}}_{v/s}(\tau)$  as the Borel transformation of $\Pi^{\textrm{OPE}}_{v/s}(q^2)$.
The coefficients $a-d$ in \eqref{sum_rule} for the  isospin $I=1$ state $1^{-+}$ state are as follows \cite{Balitsky:1986hf,Chetyrkin:2000tj,Jin:2002rw}
\begin{gather*}
a_{11}=-\frac{\alpha_s(\mu)}{240\pi^3}\left(1+\frac{1301}{240}\frac{\alpha_s(\mu)}{\pi}\right), ~~
a_{12}=\frac{\alpha_s(\mu)}{240\pi^3}\frac{17}{72}\frac{\alpha_s(\mu)}{\pi},\\
b_{11}=-\frac{1}{36\pi}\langle \alpha_s G^2\rangle \left(1-\frac{145}{72}\frac{\alpha_s(\mu)}{\pi}\right)-
\frac{2}{9}\frac{\alpha_s(\mu)}{\pi} \langle m_q\bar qq\rangle,\\
b_{12}=-\frac{1}{36\pi}\langle \alpha_s G^2\rangle \frac{8}{9} \frac{\alpha_s(\mu)}{\pi},\\
c_{11}=-\frac{4\pi}{9} \alpha_s \langle \bar qq\rangle^2\left( 1+\frac{1}{108}\frac{\alpha_s(\mu)}{\pi}\right)
-\frac{1}{192\pi^2} \langle g^3 G^3\rangle,\\
c_{12}=-\frac{4\pi}{9}\alpha_s\langle \bar qq\rangle^2\frac{47}{72}\frac{\alpha_s(\mu)}{\pi},~~
d_{11}=-\frac{1}{6}\pi\alpha_s(\mu)\langle \bar qq\rangle \langle g \bar qGq\rangle,
\end{gather*}
while for $0^{++}$ state ($I=1$), they can be written as
\begin{gather*}
a_{11}=-\frac{\alpha_s(\mu)}{480\pi^3}\left(1+\frac{6979}{2160}\frac{\alpha_s(\mu)}{\pi}\right), ~~
a_{12}=\frac{\alpha_s(\mu)}{480\pi^3}\frac{17}{72}\frac{\alpha_s(\mu)}{\pi},\\
b_{11}=\frac{1}{24\pi}\langle \alpha_s G^2\rangle \left(1-\frac{209}{72}\frac{\alpha_s(\mu)}{\pi}\right)+
\frac{\alpha_s(\mu)}{3\pi}\langle m_q \bar qq\rangle,\\
b_{12}=\frac{1}{24\pi}\langle \alpha_s G^2\rangle \frac{8}{9}\frac{\alpha_s(\mu)}{\pi},\\
c_{11}=\frac{4\pi}{3} \alpha_s \langle \bar qq\rangle^2 \left(1-\frac{37}{18}\frac{\alpha_s(\mu)}{\pi}\right)
+\frac{1}{192\pi^2}\langle g^3 G^3\rangle,\\
c_{12}=\frac{4\pi}{3} \alpha_s \langle \bar qq\rangle^2 \frac{89}{72}\frac{\alpha_s(\mu)}{\pi},~~
d_{11}=-\frac{11}{27}\pi \alpha_s(\mu) \langle \bar qq\rangle \langle g \bar qGq\rangle,
\end{gather*}
where $\alpha_s(\mu)=4\pi/(9\ln(\mu^2/\Lambda^2_{\textrm{QCD}}))$ is the running coupling constant for three flavors.

In order to obtain predictions for the ground state, the simple single narrow resonance spectral density ansatz
$\textrm{Im}\Pi^{\textrm{phen}}(s)=\pi f^2\delta(s-m^2)+\textrm{Im}\Pi^{\textrm{ESC}}(s)\theta(s-s_0)$ is typically used,
where $s_0$ is the continuum threshold that separating the contribution from excited states (ESC), $f$
and $m$  denote the coupling of the resonance to the current and the mass of the resonance respectively. Based on this assumption,
we can obtain the phenomenological representation of the correlator $\Pi^{\textrm{phen}}(\tau,s_0,f_,m)$ via the
dispersion relation \cite{Shifman:1978bx,Shifman:1978by}. The spectral density for excited states $\textrm{Im}\Pi^{\textrm{ESC}}$ can be chosen as
$\textrm{Im}\Pi^{\textrm{OPE}}(s)$, and following usual conventions, these contributions are placed on the OPE side of the sum rules.
For the present case, we obtain
\begin{equation}
\begin{split}
\Pi^{\textrm{OPE-ESC}}_{v/s}(s_0, \tau)=&a_{11}\frac{-2}{\tau^3}(1-\rho_2(s_0\tau)) +a_{12}
\left[\frac{2}{\tau^3} (2 \gamma_E-3+2\ln(\tau\mu^2))
+2 F_1(s_0,\tau)\right]\\
&+b_{11}\frac{-1}{\tau}(1-\rho_0(s_0\tau))+b_{12}\left[ \frac{2}{\tau}(\gamma_E+\ln(\tau\mu^2))+
2 F_2(s_0,\tau)\right]\\
&+c_{11}+c_{12}(-\gamma_E-\ln(\tau\mu^2)-E_1(s_0\tau))
+d_{11}\tau,
\end{split}
\end{equation}
where $\rho_n(x)=\me^{-x} \sum_{k=0}^n \frac{x^k}{k!}$, $E_1$ is the exponential integral function, and the functions
$F_1$ and $F_2$ are defined as follows
\begin{gather*}
F_1(s_0,\tau)=\frac{1}{\tau^3}[2\rho_0(s_0\tau)+\rho_1(s_0\tau)+2 E_1(s_0\tau)+2\rho_2(s_0\tau)\ln(s_0/\mu^2)],\\
F_2(s_0,\tau)=\frac{1}{\tau}[E_1(s_0\tau)+\rho_0(s_0\tau)\ln(s_0/\mu^2)].
\end{gather*}

After placing the contribution of excited states to the OPE side of the correlator, the sum rule can be written as
\begin{equation}
\label{eq:sr}
\Pi^{\textrm{OPE-ESC}}(s_0, \tau)=f^2 \me^{-m^2\tau}.
\end{equation}
This is the master equation for QCD sum rules; physical properties of relevant hadrons, i.e., $m$, $f^2$ and $s_0$,
should satisfy Eq.~\eqref{eq:sr}.

Finally, before proceeding with numerical calculations, renormalization-group (RG) improvement of the sum rules, i.e.,
substitutions $\mu^2\to1/\tau$ in Eq.\eqref{eq:sr}, is needed \cite{Narison:1981ts}. In addition, the anomalous dimensions
for condensate $\langle O\rangle$ in $\Pi^{\textrm{OPE-ESC}}$ also should
be implemented by multiplying $\langle O\rangle$
by a factor $L(\mu_0)^{\gamma_O}$, where $L(\mu_0)=[\ln(1/(\tau\Lambda^2_{\textrm{QCD}}))/\ln(\mu^2_0/\Lambda^2_{\textrm{QCD}})]$,
 $\mu_0$ is the renormalization scale for $\langle O\rangle$, and $\gamma_O$ is the anomalous dimension for condensate $\langle O\rangle$.
However, most condensates appear in Eq.~\eqref{eq:sr} are RG-invariant (e.g., $\langle \alpha_s G^2\rangle$) or approximately
RG-invariant (e.g., $\langle m_q \bar qq\rangle$, $\alpha_s\langle \bar qq\rangle^2$); the only two condensates which should
be multiplied by factor $L(\mu_0)^{\gamma_O}$ are $\langle g^3G^3\rangle$ and $\langle\bar qq\rangle\langle g\bar qGq\rangle$ whose
anomalous dimensions are -23/27 and 10/27 respectively \cite{Shifman:1978bx,book2}.
The coupling constant $f$ also should be multiplied by factor $L(m)^{\gamma_j}$, where $\gamma_j=-32/81$
to incorporate the anomalous dimension of the current \cite{Chetyrkin:2000tj,Jin:2002rw}; $f$ then receives its value at hybrid mass shell.

\section{Fitting the sum rules}

The main purpose of QCD sum rules is to obtain predictions of the lowest-lying resonance mass and coupling constant.
In traditional QCD sum rules, the prediction of resonance mass can be obtained by ratio method
\cite{Shifman:1978bx,Shifman:1978by}, i.e.,
$m^2=-\partial_\tau \ln \Pi^{\textrm{OPE-ESC}}(\tau)$,
 but this approach has some obvious shortcomings \cite{Leinweber:1995fn}, especially, fixing of $s_0$ by hand.
In order to address these shortcomings, Leinweber introduced a  Monte-Carlo based uncertainty analysis
into QCD sum rules.

Obviously, because of the truncation of OPE and the simplified assumption for the phenomenological
spectral density, Eq.~\eqref{eq:sr} is not valid for all $\tau$, thus requiring a
sum rule window in which the validity of Eq.~\eqref{eq:sr} can be established. Specifically, to ensure  convergence
of the OPE \cite{Leinweber:1995fn}, the
contributions of the highest dimensional operators (HDO) in OPE should not be too large (less than 10\% of total
OPE contributions), that will give an upper bound $\tau_{\textrm{max}}$ for the sum rule window. Meanwhile, the
continuum contributions should not be larger than the pole contributions, otherwise we can not rely on the single
narrow resonance ansatz. Demanding that the ratio of ESC/total contributions $<$ 50\% gives the lower bound $\tau_{\textrm{min}}$.
However, ESC is dependent on $s_0$, which is to be determined in QCD sum rule analysis, thus we can not determine
$\tau_{\textrm{min}}$ before doing the QCD sum rule analysis. In practice, we will  initially ``guess" a $\tau_{\textrm{min}}$,
and after finishing the analysis we can check our initial choice and adjust it iteratively until it is consistent with the results of the analysis.

The least-square method is appropriate for matching the two sides of Eq.\eqref{eq:sr} in the sum rule window. However,
the condensates appearing in OPE are not known accurately, so these parameter uncertainties will lead to uncertainties in the OPE. Because
the uncertainties in the OPE are not equal at different points in the sum rule window, a weighted-least-square method
is more appropriate.

Following Leinweber's procedure, we first need to estimate the standard deviation $\sigma_{\textrm{OPE}}(\tau)$ of
$\Pi^{\textrm{OPE}}(\tau)$ at any point $\tau$ in the sum rule window. This can be done 
by randomly generating 200 set of Gaussian distributed input parameters (condensates
and $\Lambda_{\textrm{QCD}}$) with given uncertainties. 200 samples are enough to establish  the stable standard deviation. 
After obtaining $\sigma_{\textrm{OPE}}(\tau)$, the phenomenological output parameters $s_0$, $f^2$ and $m$ can be obtained
by minimizing a weighted $\chi^2$, which is defined as follows
\begin{equation}
\chi^2=\sum_{j=1}^{n_B}\frac{(\Pi^{\textrm{OPE}}(\tau_j)-\Pi^{\textrm{phen}}(\tau_j,s_0,f,m))^2}{\sigma_{\textrm{OPE}}^2(\tau_j)}.
\end{equation}
The points $\tau_j$ are selected to be $\tau_j=\tau_\textrm{min}+(\tau_\textrm{max}-\tau_\textrm{min})\times(j-1)/(n_B-1)$,
i.e., we divide the sum rule window into ($n_B-1$) evenly parts. In the $1^{-+}$ light hybrid case, we find the fitting results do
not change provided $n_B\geq 8$, thus we set $n_B=21$ for simplicity.

In this paper, we generate a set of 2000 Gaussian distributed input parameters with given uncertainties, and for each set
we minimize $\chi^2$ to obtain a set of fitted phenomenological output parameters. Finally, we will select the physical
results (it is natural that there exist constraints
between output parameters, such as $s_0>m^2$ etc, and results that violate these constraints should be excluded) from the
set of fitted values. An uncertainty analysis of phenomenological output parameters is then possible.
In addition,  scatter plots of input  and output parameters allow us to study correlations
between them, and these studies can shed some light on how to improve the accuracy of the analysis.

\section{Numerical results for $1^{-+}$ light hybrid}

To randomly generate Gaussian distribution input parameters, we first set the central values and uncertainties of these parameters.
In this paper, we treat all condensates as independent parameters in order to distinguish the importance of different condensates.
After reviewing the literature, we choose the central value of input parameters at $\mu_0=1$\,GeV as follows
\cite{Leinweber:1995fn,Jin:2000ek,Jin:2002rw}
\begin{gather*}
\Lambda_{\textrm{QCD}}=0.20\,\textrm{GeV},~~\langle \alpha_s G^2\rangle =0.095\,\textrm{GeV}^4,~~m_q\langle \bar qq\rangle
=0.007\times(-0.236)^3\,\textrm{GeV}^4,\\
\langle g^3 G^3\rangle =1.1\times 0.095\,\textrm{GeV}^6,~~
\alpha_s\langle\bar qq\rangle^2=1.8\times10^{-4}\,\textrm{GeV}^4,~~
\langle \bar qq\rangle\langle g\bar qGq\rangle =(-0.236)^6\times0.72\,\textrm{GeV}^8.
\end{gather*}
Then all sets of input parameters are generated with 10\% uncertainties, which is a typical uncertainty in QCDSR.
For physical considerations, we add an additional constraint $0.10\,\textrm{GeV}\leq\Lambda_{\textrm{QCD}}\leq
0.30\,\textrm{GeV}$ on $\Lambda_{\textrm{QCD}}$. Any set of randomly generated input parameters which violates this constraint
will be excluded from our set of input parameters.

\begin{figure}[htbp]
\centering
\includegraphics[scale=0.9]{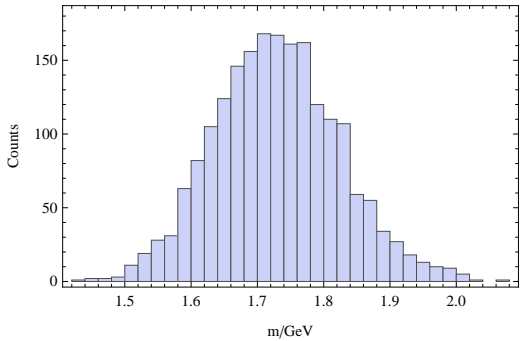}
\caption{\label{fig:mhistogram} The histogram of $1^{-+}$ light hybrid masses obtained from 2000 matches.}
\end{figure}

After several numerical samples, we find the appropriate sum rule window for the $1^{-+}$ light hybrid is $\tau=0.4-1.0\,\textrm{GeV}^{-2}$.
By minimizing the $\chi^2$ function for each sample of input parameters, we finally obtain a 2000-member set of phenomenological output parameters
which satisfy our physical constraints.
In Fig.\ref{fig:mhistogram}, we plot the histogram for these 2000 different $1^{-+}$ light hybrid masses  obtained in the least-squares fitting procedure. It is obvious that the distribution of $m$ is very close to a Gaussian.

\begin{figure}[htbp]
\centering
\includegraphics[scale=0.9]{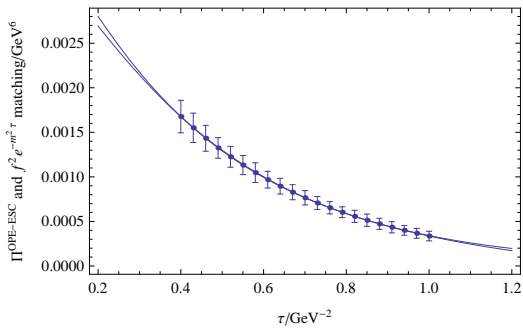}
\caption{\label{fig:fit} The least-squares fit of the $1^{-+}$ light hybrid with the median value for $f^2$, $m$, $s_0$ and all condensates.
The error bars show $\sigma_{\textrm{OPE}}(\tau_j)$ estimated at each $\tau_j$, and the factor $L(m)^{\gamma_j}$ is absorbed
in $f$.}
\end{figure}

The main results of our fitting procedure are as follows
\begin{equation}
s_0=5.37^{+0.81}_{-0.62}\,\textrm{GeV}^2,~~m=1.73^{+0.10}_{-0.09}\,\textrm{GeV},~~f^2=0.0053^{+0.0013}_{-0.0009}\,\textrm{GeV}^6,
\end{equation}
where we have reported the median and the asymmetric standard deviations from the median \cite{book} for all physical output parameters.

\begin{figure}[htbp]
\centering
\includegraphics[scale=0.9]{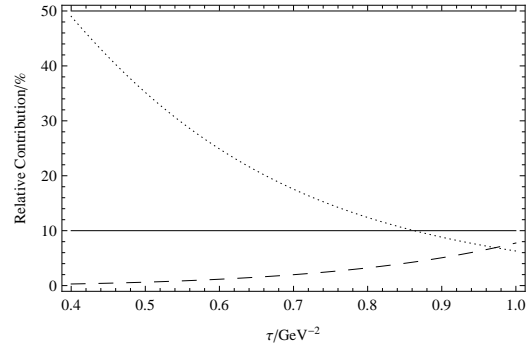}
\caption{\label{fig:srwin} Sum rule window for the $1^{-+}$ light hybrid. The dotted line denotes the relative excited states contributions while
the dashed line denotes the relative HDO contributions.}
\end{figure}

The uncertainty of $m$ is less than 6\%, implying that the fitted results are very stable with different input parameters.
Fig.\ref{fig:fit} illustrates that the fits are qualitatively acceptable,
and from Fig.\ref{fig:srwin} we demonstrate that the
two conditions for determining the sum rule window are met very well for the median value of phenomenological parameters.

\begin{figure}[htbp]
\centering
\includegraphics[scale=0.9]{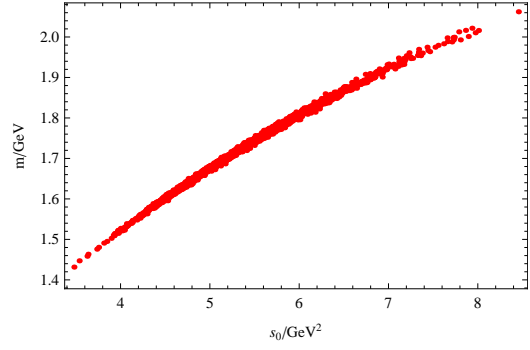}
\caption{\label{fig:s0m} The scatter plot of $1^{-+}$ light hybrid mass and the continuum threshold.}
\end{figure}

\begin{figure}[htbp]
\centering
\includegraphics[scale=0.9]{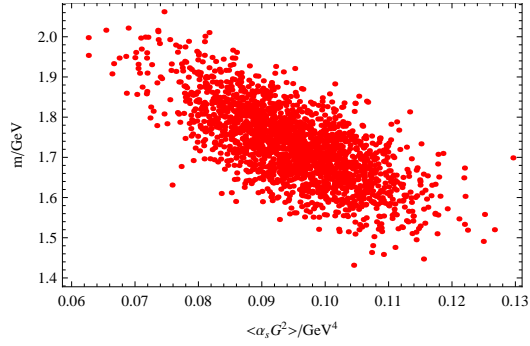}
\caption{\label{fig:magg} The scatter plot of $1^{-+}$ light hybrid mass and two-gluon condensate.}
\end{figure}

\begin{figure}[htbp]
\centering
\includegraphics[scale=0.9]{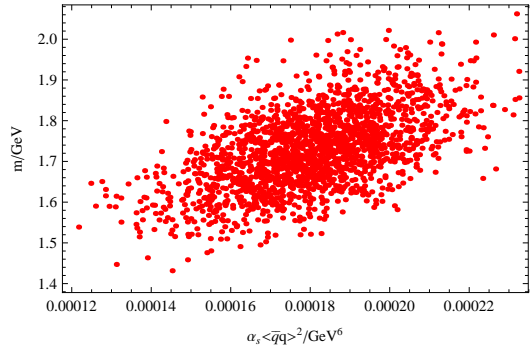}
\caption{\label{fig:mqqqq} The scatter plot of $1^{-+}$ light hybrid mass and four-quark condensate.}
\end{figure}

\begin{figure}[htbp]
\centering
\includegraphics[scale=0.9]{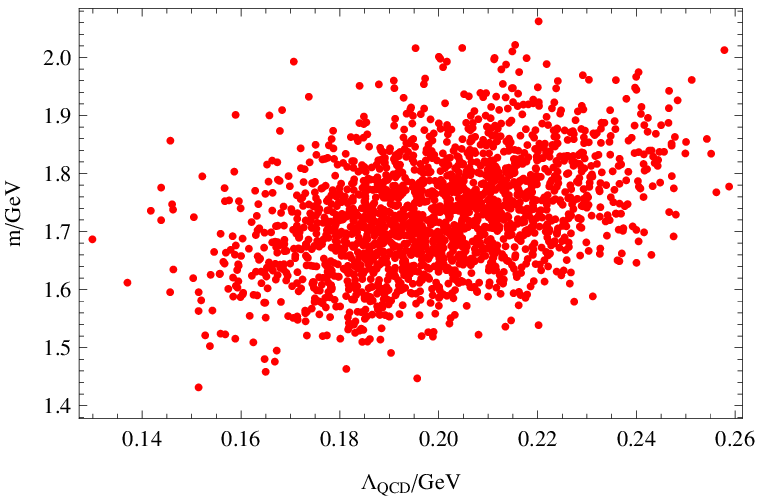}
\caption{\label{fig:mlambda} The scatter plot of $1^{-+}$ light hybrid mass and $\Lambda_{\textrm{QCD}}$.}
\end{figure}

Based on the results of our least-squares fits, the correlations of the input and output parameters can also be obtained.
From the scatter plot Fig.~\ref{fig:s0m} we find that there exists a very strong positive correlation between  $m$ and $s_0$ that
may explain why a broad range of $m$ can occur in traditional sum rule analyses by setting an appropriate $s_0$ \cite{Matheus:2004qq}.
From Fig.~\ref{fig:magg} we find that there exists a weak negative correlation between $m$ and $\langle \alpha_s G^2\rangle$;
the Wilson coefficient and the value for $\langle\alpha_s G^2\rangle$ are thus crucial to our results. Further study should improve
the present understanding on this point. Furthermore, we find there exists weak positive correlations between $m$ and four-quark condensate,
and for $\Lambda_{\textrm{QCD}}$ (see Figs.~\ref{fig:mqqqq} and \ref{fig:mlambda}). Thus the Wilson coefficient and possible
factorization violation effects for the four-quark condensate also play an important role in the result. It is somewhat
unexpected that the present result is sensitive to $\Lambda_{\textrm{QCD}}$ in contrast to the $\rho$ meson case \cite{Leinweber:1995fn}.
The possible explanation may come from the fact that the leading term of OPE is $\alpha_s$-dependent  and  the anomalous dimension is not zero  in the current.

\begin{table}[htbp]
\caption{\label{tab:result} Matching results with larger uncertainties for input parameters. All matching results with
$\Lambda_{\textrm{QCD}}>0.30\,\textrm{GeV}$ or  $\Lambda_{\textrm{QCD}}<0.10\,\textrm{GeV}$ are excluded.}
\begin{ruledtabular}
\begin{tabular}{cccccc}
  Parameters with 30\% uncertainties\footnotemark[1]  & $\langle \alpha_s G^2\rangle$ & $\alpha_s \langle \bar qq\rangle^2$ &
  $\Lambda_{\textrm{QCD}}$ & $\langle \alpha_s G^2\rangle$, $\alpha_s \langle \bar qq\rangle^2$, $\Lambda_{\textrm{QCD}}$ & all parameters\\
  \hline
  Output $s_0$/GeV$^2$ & $5.24^{+1.83}_{-1.03}$ & $5.26^{+1.51}_{-1.10}$ &  $5.50^{+1.07}_{-0.89}$ & $5.26^{+1.96}_{-1.40}$ & $5.28^{+2.01}_{-1.45}$ \\
  Output $m$/GeV & $1.71^{+0.21}_{-0.16}$ & $1.72^{+0.17}_{-0.17}$ & $1.73^{+0.13}_{-0.13}$ & $1.71^{+0.22}_{-0.22}$ & $1.71^{+0.22}_{-0.23}$ \\
  Output $f^2$/GeV$^6$ & $0.0052^{+0.0027}_{-0.0011}$ & $0.0052^{+0.0028}_{-0.0016}$ & $0.0055^{+0.0023}_{-0.0018}$ &
  $0.0053^{+0.0035}_{-0.0018}$ & $0.0053^{+0.0035}_{-0.0019}$\\
\end{tabular}
\end{ruledtabular}
\footnotetext[1]{Other input parameters are still generated with 10\% uncertainties.}
\end{table}

We may also want to estimate how the result will change if the uncertainties for input parameters are larger than 10\%.
Table~\ref{tab:result} shows that the uncertainties for $\langle\alpha_s G^2\rangle$, $\alpha_s\langle \bar qq\rangle^2$
and $\Lambda_{\textrm{QCD}}$ play the most important roles in the least-squares fit results, i.e., the uncertainties
of output parameters are mainly determined by these three input parameters. Table~\ref{tab:result} also
demonstrates that the medians of $f^2$ and $m$ change a little while $s_0$ is somewhat sensitive to larger uncertainties for input
parameters. It probably  means that the continuum absorbs these uncertainties in some way.

Finally, if we want to improve the simplified single narrow resonance spectral density, we may use a Breit-Wigner form
($\frac{1}{\pi}m \Gamma/((s-m^2)^2 +m^2 \Gamma^2)$) spectral density instead, this modification
 replaces $\me^{-m^2\tau}$ in Eq.~\eqref{eq:sr} with
\begin{equation*}
\frac{1}{\pi}{\rm Im}\left[\me^{-m^2 \tau+\mi\, m\Gamma\tau}{\rm Ei}(m^2\tau-\mi\, m \Gamma \tau)\right]
-\frac{1}{\pi}{\rm Im}\left[\me^{-m^2 \tau+\mi\, m\Gamma\tau}{\rm Ei}(m^2\tau-s_0\tau-\mi\, m\Gamma\tau)\right],
\end{equation*}
where $\Gamma$ is the width of the resonance and Ei is an exponential integral function. However, a four parameter fit ($m$,
$f^2$, $s_0$ and $\Gamma$) search always give the same result as the three parameter fit ($m$, $f^2$ and $s_0$), i.e.,
gives a result with $\Gamma=0$ automatically. If we set a ``reasonable"  non-zero $\Gamma$ as input, then a three parameter fit ($m$,
$f^2$ and $s_0$) will give a result with a slightly increased value for the predicted mass. However, in the present case, the
weighted $\chi^2$ will increase significantly, implying that the goodness of fit with non-zero $\Gamma$ is worse than with $\Gamma=0$.
This situation not only occurs in the hybrid,  we have checked that it also occurs for the $\rho$ meson.
These specific results are in agreement with the general argument presented in Ref.~\cite{Elias:1998bq}.

\section{Numerical result for $0^{++}$ light hybrid}

The procedure of $0^{++}$ light hybrid does not give a result as good as the $1^{-+}$ case. Because of the sign of the
$\langle\alpha_s G^2\rangle$ in OPE, it is more difficult to find an appropriate sum rule window for $0^{++}$
case. We can only find a small sum-rule region of validity, $\tau =0.08-0.29\,\textrm{GeV}^{-2}$, for the $0^{++}$ case. The small sum rule
window may reduce the reliability of the least-squares fit result. Because the $0^{++}$ quantum numbers are non-exotic, it is possible that mixing of the hybrid with $\bar q q$ and gluonic currents is needed to study the scalar channels.  Based on mixing of hybrid charmonium and molecular systems, the large mass of this  $0^{++}$ hybrid  does not preclude the possibility of a mixed interpretation \cite{Chen:2013pya}.

We also notice that the uncertainties of phenomenological output parameters (obtained by generating input parameters with 10\%
uncertainties)
\begin{equation}
s_0=38.9^{+11.2}_{-9.0}\,\textrm{GeV}^2,~~m=4.86^{+0.34}_{-0.45}\,\textrm{GeV},~~f^2=0.22^{+0.15}_{-0.10}\,\textrm{GeV}^6,
\end{equation}
are larger than in the $1^{-+}$ hybrid case. The correlations also are different with the $1^{-+}$ case; there now exist positive
correlations between $m$ and $\langle \alpha_sG^2\rangle$, and negative correlations between $m$ and $\Lambda_{\textrm{QCD}}$.
The $\alpha_s\langle \bar qq\rangle^2$ condensate is not important in the present case.

\section{Discussion and Summary }

In this paper, we have reanalyzed the $1^{-+}$ and $0^{++}$ light hybrids from QCD sum rules with a Monte-Carlo based uncertainty analysis.
In this procedure, the continuum threshold $s_0$ now is an output rather than an input parameter, and thus we avoid some subjective factors.
Based on the uncertainties of our least-squares fit results, we conclude that a reliable mass prediction has been obtained for the $1^{-+}$ light
hybrid mass. Input parameters with 10\% uncertainties cause about only 6\% uncertainty in the mass, while 30\% uncertainties cause about
13\% uncertainty in the mass. We choose the latter as a cautious estimate, thus we predict the mass  of the $1^{-+}$ light hybrid
is $1.71\pm 0.22$\,GeV. This result favors  $\pi_1(1600)$ to be a hybrid, but  $\pi_1(1400)$ is not completely excluded
because some input parameters may have larger uncertainties.
The uncertainty of  $f^2$ seems much larger, but it could be easily understood by re-parameterizing $f=m^2 f^\prime$.
Then, the uncertainty of new decay constant $f^\prime$ is as small as the uncertainty of $m$
 (input parameters with 10\% uncertainties cause about 6\% uncertainty in $f^{\prime 2}$).
  We also investigate  the sum rules with a Breit-Wigner form spectral density.  We find the best fit to the sum rules
 gives a result with $\Gamma=0$. It is really a surprising result but it is not in isolation; we have checked it is also true for the $\rho$
meson. Obviously, the procedure is not sensitive to the decay width at present. How to obtain a non-zero decay width needs to be further studied.
In principle, this method can be extended to study excited
states, although some new assumptions are necessary. For instance, if we assume $s_0=(m^2+m^2_1)/2$, where $m_1$ is the next
excited state \cite{Kataev:1982et,Kataev:1982xu,Krasnikov:1981vw}, then the predicted  mass of the first excited
$1^{-+}$ hybrid is  about 3\,GeV.

The uncertainties of output parameters for the $0^{++}$ light hybrid are larger than the $1^{-+}$ case, thus the results are less
reliable. According to our result, if $0^{++}$ pure light hybrid state exists, its mass will be heavier than any known $a_0$ meson.
Further experiment or mixted scenarios containing a hybrid component may address this question.

In conclusion,  the QCD sum rule uncertainties  arise not only from OPE truncation and the single resonance assumption, but also from
sensitivity to the uncertainties of input parameters. So any corrections, higher order $\alpha_s$ to the coefficients of the
condensates or higher dimension contributions would be valuable.
By contrast,  the decay width of the  light hybrid is below the sensitivity of the present method,
and hence the QCD sum-rules analysis is not dependent on the form of the spectral density.


\begin{acknowledgments}
This work is partly supported by K. C. Wong Magna Fund in Ningbo University and NSFC under grant 11205093 and 11175153.
TGS is grateful for research support from the Natural Sciences and Engineering Research Council of Canada (NSERC).
\end{acknowledgments}


\end{document}